\documentclass[pra,twocolumn,showpacs,preprintnumbers,amsmath,amssymb,aps]{revtex4}
\usepackage{color}
\usepackage{bm, graphicx, amsmath}
\usepackage{hyperref}

\usepackage{braket}

\begin{document}
\title{Broadcast of a restricted set of qubit and qutrit states}
\author{Mark Hillery$^{1,2}$, J\'{a}nos A. Bergou$^{1,2}$, Tzu-Chieh Wei$^{3}$, Siddhartha Santra$^{4}$, and Vladimir Malinovsky$^{4}$}
\affiliation{$^{1}$Department of Physics and Astronomy, Hunter College of the City University of New York, 695 Park Avenue, New York, NY 10065 \\
$^{2}$Graduate Center of the City University of New York, 365 Fifth Avenue, New York, NY 10016 \\
$^{3}$C.\ N.\ Yang Institute for Theoretical Physics and Department of Physics and Astronomy, State University of New York at Stony Brook, NY 11794-3840 \\
$^{4}$U. S. Army Combat Capabilities Development Command, Army Research Laboratory, Adelphi, MD 20783} 
\begin{abstract}
The no-cloning theorem forbids the distribution of an unknown state to more than one receiver.  However, if the sender knows the state, and the state is chosen from a restricted set of possibilities, a procedure known as remote state preparation can be used to broadcast a state.  Here we examine a remote state preparation protocol that can be used to send the state of a qubit, confined to the equator of the Bloch sphere, to an arbitrary number of receivers.  The entanglement cost is less than that of using teleportation to accomplish the same task.  We present a number of variations on this task, probabilistically sending an unknown qubit state to two receivers, sending different qubit states to two receivers, and sending qutrit states to two receivers.  Finally, we discuss some applications of these protocols.
\end{abstract}

\maketitle

\section{Introduction}
Quantum networks are multiple-user systems for the transmission and routing of quantum information.   Qubits are exchanged between users either directly or by means of teleportation.  In the latter case, the parties need to share entangled states, and the distribution of this entanglement is one of the chief tasks in operating the network. 

In some cases a user would like to broadcast the same quantum information to several users.  In this case the no-cloning theorem can come into play.  If the quantum state to be transmitted is unknown, then one can use a quantum cloner to make imperfect copies of the state \cite{BH}, and send these to the desired parties.  This may not be satisfactory, however, exactly because the clones are imperfect.  If one wants to distribute a known state, an alternate procedure known as remote state preparation, which requires less classical communication than teleportation, can be used \cite{Lo,Bennett,Leung,Hayden,Zukowski}.  A number of schemes for broadcasting qubit states have been investigated, but they typically require $N$ ebits if a state is broadcast to $N$ parties \cite{Yu1,Yu2,Zhou,Zhao}.   

The situation becomes simpler if one wants to send a restricted class of states \cite{Lo,pati}.  For example, in the case of sending a state to just one party, let us suppose Alice wants to send one of the states $|\psi (\theta )\rangle = (e^{i\theta}|0\rangle + e^{-i\theta} |1\rangle )/\sqrt{2}$ , where $0 \leq \theta < \pi$, to Bob, and that they share a singlet, $(|0\rangle_{a}|1\rangle_{b} - |1\rangle_{a}|0\rangle_{b})/\sqrt{2}$.  Alice measures her part of the singlet in the $\{ |\psi (\theta )\rangle , |\psi (\theta+ \pi /2)\rangle \}$ basis, and tells Bob her result.  If she gets $|\psi (\theta +\pi /2)\rangle$, Bob does nothing, and if she gets $|\psi (\theta )\rangle$, then Bob applies the operator $\sigma_{z}$ to his state.  Note that this procedure is cheaper in terms of classical communication than is teleportation.  It requires one bit of classical communication whereas teleportation requires two, because of the special form of the equatorial state~\cite{Graham2015}.  Remote state preparation can be adapted to send states to more than one party \cite{pati}.  The protocol proposed by Agrawal, \emph{et\ al.,} in \cite{pati}, makes use of dark states, and if a qubit state is to be distributed to $N$ parties, an entangled state of $2N$ qubits is required, for an entanglement cost of $N$ ebits.

Here we wish to present a procedure for distributing a limited class of states that is less costly in terms of entanglement and classical communication.  It makes use of a qudit for the sender, and one qubit for each receiver.  This scheme can be modified to send different states to different parties, at the cost of more entanglement, or to send qutrits instead of qubits.

\section{Procedure}
\subsection{Basic protocol}
Let us begin with a simple situation.  Alice wants to send the state $e^{i\theta} \alpha |0\rangle + e^{-i\theta}\beta |1\rangle$ to Bob and Charlie, where $\alpha$ and $\beta$ are fixed, and Alice can control $\theta$.  She starts with two qubits in the state $(\alpha |0\rangle_{b} + \beta |1\rangle_{b})(\alpha |0\rangle_{c} + \beta |1\rangle_{c})$ and entangles them with a qutrit, see Fig.\ 1, to form the state
\begin{eqnarray}
\label{template}
|\Psi\rangle_{abc} & = & \alpha^{2}|0\rangle_{a} |00\rangle_{bc} + \beta^{2} |1\rangle_{a} |11\rangle_{bc} \nonumber \\
& & + \alpha\beta |2\rangle_{a} (|01\rangle_{bc} + |10\rangle_{bc} ) .
\end{eqnarray}
Note that the normalization condition for the this state,  $|\alpha|^{2}+|\beta|^{2}=1$, is the same as for the qubit states Alice wants to send.  She now sends qubit $b$ to Bob and qubit $c$ to Charlie.  After determining what value of $\theta$ she wants to send, she applies the operator $U_{a}$ to her qutrit, where $U_{a}|0\rangle_{a} = e^{2i\theta}|0\rangle_{a}$, $U_{a}|1\rangle_{a} = e^{-2i\theta}|1\rangle_{a}$, and $U_{a}|2\rangle_{a} = |2\rangle_{a}$.  She now measures her qutrit in the basis
\begin{eqnarray}
|u_{0}\rangle_{a} & = & \frac{1}{\sqrt{3}}( |0\rangle_{a} + |1\rangle_{a} + |2\rangle_{a}) \nonumber \\
|u_{1}\rangle_{a} & = & \frac{1}{\sqrt{3}}( e^{2\pi i/3}|0\rangle_{a} + e^{-2\pi i/3}|1\rangle_{a} + |2\rangle_{a}) \nonumber \\
|u_{0}\rangle_{a} & = & \frac{1}{\sqrt{3}}( e^{-2\pi i/3}|0\rangle_{a} + e^{2\pi i/3}|1\rangle_{a} + |2\rangle_{a}) 
\end{eqnarray}
If she gets $|u_{0}\rangle_{a}$, then Bob and Charlie each get $e^{i\theta} \alpha |0\rangle + e^{-i\theta}\beta |1\rangle$, and the procedure has been successful.  If she gets $|u_{1}\rangle_{a}$, Bob and Charlie get the desired state after they have both applied the operator $U_{b}$ to their qubits, where $U_{b}|0\rangle = e^{i\pi /3} |0\rangle$ and $U_{b}|1\rangle = e^{-i\pi /3} |1\rangle$.  If Alice gets $|u_{2}\rangle$, Bob and Charlie get the desired state if they each apply $U_{b}^{-1}$ to their qubits.  Therefore, Bob and Charlie get the desired state with probability one.  Note that all of the operations and measurements are local.   

\begin{figure}
\begin{center}
\includegraphics[width=24em]{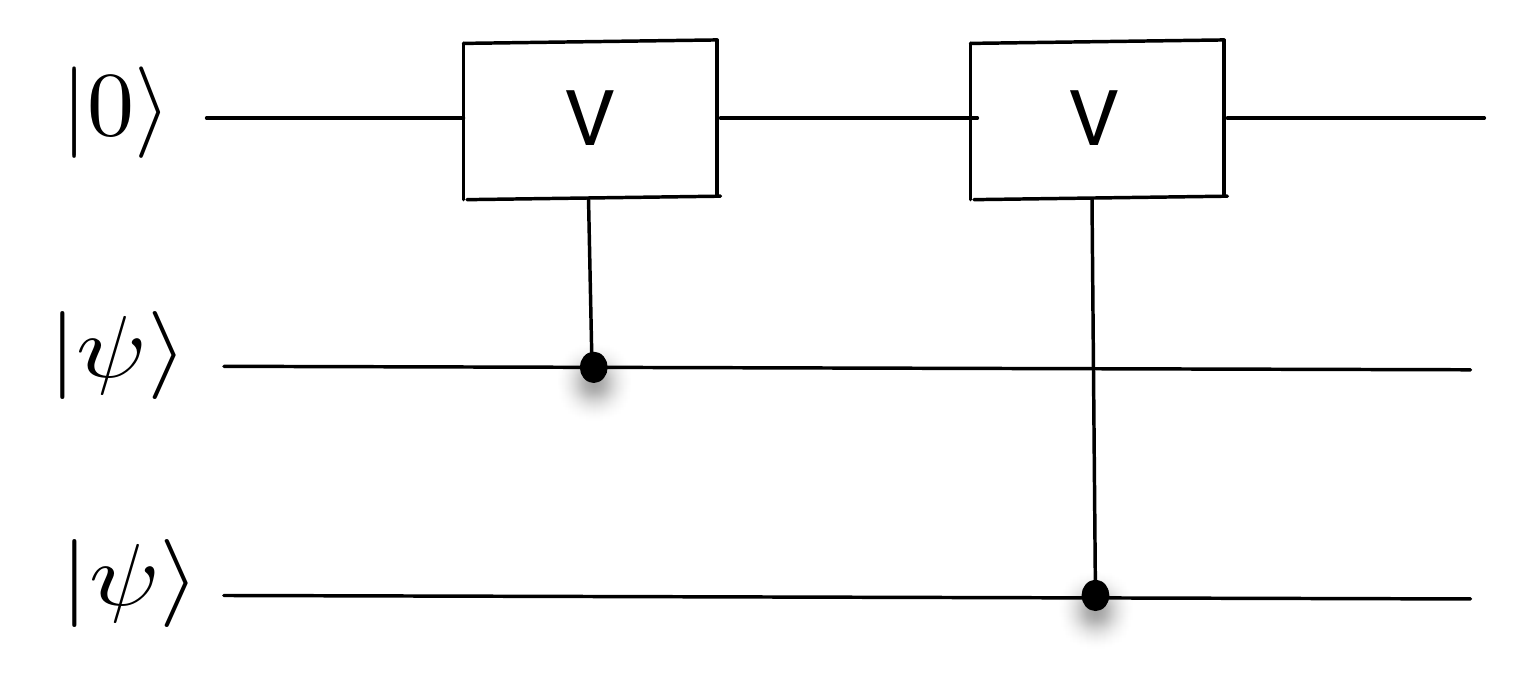}
\end{center}
\caption{Circuit for producing the state in  Eq.\ (\ref{template}).  The top line is a qutrit, initially in the state $|0\rangle$, and the bottom two lines are qubits, both initially in the state $|\psi\rangle = \alpha |0\rangle + \beta |1\rangle$.}
\end{figure}

Note that Alice can create the state in Eq.\ (\ref{template}) and distribute the qubits at one time and perform the operation $U_{a}$ and her measurement at a later time.  The entangled state resource can be distributed and stored until it is needed.  The state can be created using two Controlled-V gates (see Fig.\ 1) , where $V$ is the operator acting on the qutrit, $V|j\rangle = |j-1\rangle$, where $j=0,1,2$ and the addition is modulo $3$.  The control line is a qubit, and if it is in the state $|0\rangle$, the gate does nothing to the target qutrit, but if the qubit is in the state $|1\rangle$, $V$ is applied to the qutrit.  The circuit consists of three lines, with the qutrit in line $1$ and the qubits in lines $2$ and $3$.  There is a Controlled-V gate between lines $1$ and $2$ and a second between lines $1$ and $3$.  If the initial state of the circuit is $|0\rangle_{1} (\alpha |0\rangle_{2} + \beta |1\rangle_{2})  (\alpha |0\rangle_{3} + \beta |1\rangle_{3})$, the output is the state in Eq.\ (\ref{template}).

The same result, distributing the state $e^{i\theta} \alpha |0\rangle + e^{-i\theta}\beta |1\rangle$ to Bob and Charlie, can be accomplished using teleportation, Alice could teleport the desired state to both Bob and Charlie, but that would require two (Bell-state) measurements, while the above procedure requires one measurement that does not require projecting onto an entangled basis.  In addition, the entanglement required between Alice, on the one hand, and Bob and Charlie, on the other, is $2$ ebits in the case of teleportation and a maximum of less than $\log_{2}3$ for the above procedure.   

\begin{figure}[ht]
\begin{center}
\includegraphics[width=24em]{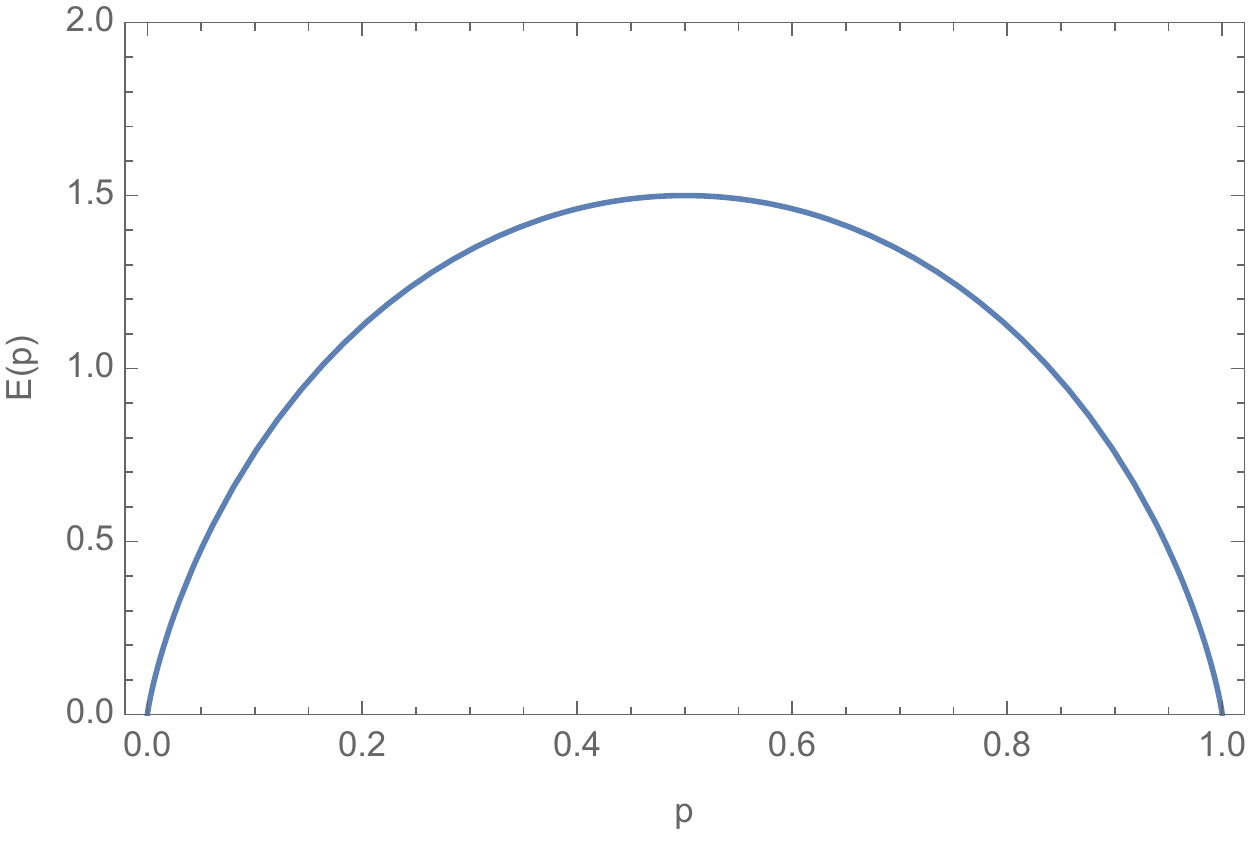}
\end{center}
\caption{The entanglement entropy, $E(p)$ from Eq. \eqref{E}, vs. $p$. The entropy has one maximum, at $p_{0}=0.5$. The maximum value at the extremal point is $E(p_{0})=1.5$.}
\label{Fig2}
\end{figure}

To be more precise, we are comparing two scenarios.  In the first, Alice prepares the state in Eq.\ (\ref{template}) and sends the qubits to Bob and Charlie.  The entanglement, calculated by tracing out Bob and Charlie and finding the von Neumann entropy of Alice's reduced density matrix, is 
\begin{eqnarray}
-|\alpha |^{4} \log_{2}|\alpha |^{4} - |\beta |^{4} \log_{2}|\beta |^{4} -2|\alpha\beta |^{2} \log_{2} (2|\alpha\beta |^{2}) \nonumber \\
\leq \log_{2}3 ,
\end{eqnarray}
but numerically a sharper bound can be found. Let us introduce the notation $p\equiv |\alpha|^{2}$. Then the above equation can be cast to the equivalent but simpler form,
\begin{eqnarray}
E(p) = - 2p \log_{2}p - 2(1-p) \log_{2}(1-p) - 2p(1-p) .
\label{E}
\end{eqnarray}
Thus, the entanglement entropy $E$ depends on the single parameter $p$. It is plotted in Fig. \ref{Fig2}. From the figure it is obvious that  $E(p) \leq  1.5 < \log_{2}3$.

In the second scenario, Alice creates two singlet states, and sends one qubit from each pair to Bob and Charlie.  In this case, the entanglement between Alice on the one hand and Bob and Charlie on the other, which is given by the von Neumann entropy of Alice's reduced density matrix, is $2$.  Finally, in the case of teleportation, the classical information content of the messages Alice sends to Bob and Charlie is $4$ bits, two bits to Bob and two bits to Charlie, while in the case of this remote state preparation protocol it is $\log_{2}3$ bits, since Bob's and Charlie's messages are the same, equiprobable, and there are three possible messages.

\subsection{Variations}
The basic protocol can be modified, and here we present two such variations.  One allows Bob and Charlie to use different bases, and the second allows Alice to probabilistically send an unknown angle.

It is possible to send states in different bases to Bob and Charlie.  Suppose Alice wants to send $(1/\sqrt{2}) ( e^{i\theta} |0\rangle + e^{-i\theta}|1\rangle )$ to Bob and $(1/\sqrt{2}) (e^{i\theta} |+x\rangle + e^{-i\theta}|-x\rangle )$ to Charlie, where $|\pm x\rangle = (1/\sqrt{2}) (|0\rangle \pm |1\rangle )$.  Alice starts with the state
\begin{eqnarray}
\frac{1}{2} [ |0\rangle_{a} |0\rangle_{b} |+x\rangle_{c} + |1\rangle_{a} |1\rangle_{b} |-x\rangle_{c} \nonumber \\
+ |2\rangle_{a} (|0\rangle_{b}|-x\rangle_{c} + |1\rangle_{b} |+x\rangle_{c} ) ]  .
\end{eqnarray}
Note that this state can be created from the state in Eq.\ (\ref{template}), with $\alpha = \beta = 1/\sqrt{2}$, by rotating qubit $c$.
She sends qubit $b$ to Bob and qubit $c$ to Charlie, and applies $U_{a}$ to her qutrit as before.  She measures her qutrit in the same basis as previously, and announces her result.  In order to perform the correction operations, Bob uses the same unitary operator as before, but Charlie now uses $U_{c}$ instead, where $U_{c}|\pm x\rangle_{c}= e^{\pm i\pi /3} |\pm x\rangle_{c}$.  If Alice gets $|u_{0}\rangle_{a}$, then they do nothing, if she gets $|u_{1}\rangle_{a}$, Bob applies $U_{b}$ and Charlie applies $U_{c}$, and if she gets $|u_{2}\rangle_{a}$, then Bob applies $U_{b}^{-1}$ and Charlie applies $U_{c}^{-1}$.

Alice can also send states, with an angle unknown to her, to Bob and Charlie probabilistically, that is, the transmission will succeed with a probability less than one.  Suppose she has distributed qubits in the state in Eq.\ (\ref{template}) to Bob and Charlie, and she receives the qutrit state
\begin{equation}
|\Phi\rangle_{d} = \frac{1}{\sqrt{3}} (e^{2i\theta} |0\rangle_{d} + |1\rangle_{d} + e^{-2i\theta} |2\rangle_{d})  ,
\end{equation}
which encodes the angle $\theta$.  She sends the two qutrits through a controlled-W ($C-W$) gate, with the $a$ qutrit as the control and the $d$ qutrit as the target.  The operation $W$ acts on a qutrit as $W|j\rangle = |j+1\rangle$, where the addition is modulo $3$.  The $C-W$ gate acts as
\begin{equation}
(C-W)_{ad} = |0\rangle_{a}\langle 0|\otimes I_{d} + |1\rangle_{a}\langle 1|\otimes W_{d}  + |2\rangle_{a}\langle 2|\otimes W^{2}_{d} .
\end{equation}
Alice then measures the $d$ qutrit in the $\{ |j\rangle_{d}\, |\, j=0,1,2\}$ basis, and if she obtains $|0\rangle_{d}$, which she does with a probability of $1/3$, then the resulting $abc$ state is now just $U_{a}|\Psi\rangle_{abc}$, and the protocol proceeds as before. Because the angle $\theta$ is unknown, the correction step as in the basic protocol is not possible.

\section{N qubits}
This procedure can be extended to $N$ parties (we will consider only the case in which all parties are using the same basis).  Alice now prepares a state consisting of one qudit with $N+1$ levels and $N$ qubits 
\begin{equation}
|\Psi\rangle = \sum_{k=0}^{N} \alpha^{k} \beta^{N-k} \left( \begin{array} {c} N \\ k \end{array}\right)^{1/2} |k\rangle_{a} |k;N-k\rangle .
\end{equation}
Here $|k\rangle_{a}$ is the state of the qudit, and $|k;N-k\rangle$ is an $N$-qubit state, which is a normalized completely symmetric state in which $k$ of the qubits are in the state $|0\rangle$ and $N-k$ are in the state $|1\rangle$.  These $N$-qubit states are known as Dicke states.  This state can be prepared by starting the qudit in the state $|0\rangle$ and the qubits in the state $|+x\rangle$.  Each qubit is connected to the qudit by a Controlled-Shift gate, where the qubit is the control and the qudit the target.  The gate acts as (qubit state first) $|0\rangle |j\rangle \rightarrow |0\rangle |j\rangle$ and $|1\rangle |j\rangle\rightarrow |1\rangle |j+1\rangle$, where the addition is modulo $N+1$.
From the perspective of angular momentum, the Dicke states are angular-momentum eigenstates, with $S_{\rm tot}=N/2$ and $S_z=k/2$, and are in the highest-weight sector via adding angular momenta of $N$ qubits. The collective effect of  Controlled-Shift gates by the $N$ qubits on the qudit state is to shift the latter in the qudit basis by an amount equal to the value of $2S_z=k$ (or shifting its z-component angular momentum by $S_z=k/2$).

Alice distributes the qubits to the $N$ parties.  She can now choose the angle by applying the operator $U_{a}$ to the qudit, where
\begin{equation}
U_{a}(\theta )|k\rangle_{a} = e^{i(2k-N)\theta }|k\rangle_{a} .
\end{equation}
Alice now  measures the qudit in the basis
\begin{equation}
|u_{m}\rangle = \frac{1}{\sqrt{N+1}} \sum_{k=0}^{N} e^{2\pi imk/(N+1)} |k\rangle_{a} ,
\end{equation}
where $m=0,1, \ldots N$.  If she obtains the result $|u_{m}\rangle$ the unnormalized $N$ qubit state is
\begin{eqnarray}
|\Phi_{m}\rangle & = & \frac{1}{\sqrt{N+1}} \sum_{k=0}^{N} \alpha^{k} \beta^{N-k} \left( \begin{array} {c} N \\ k \end{array}\right)^{1/2} e^{i(2k-N)\theta} \nonumber \\
& &  e^{-2\pi ikm/(N+1)} |k\rangle_{a} |k;N-k\rangle .
\end{eqnarray}
Alice then broadcasts the result of her measurement, and each of the parties applies the correction operator, $U_{m}$ to their qubit, where $U_{m}|0\rangle = \exp [2\pi im/(N+1)] |0\rangle$ and $U_{m}|1\rangle = |1\rangle$.  This will result in each party possessing the state $\alpha e^{i\theta}|0\rangle + \beta e^{-i\theta}|1\rangle$.

Let us compare this procedure to teleporting the states to the parties.  Teleportation would require $N$ Bell state measurements and an entanglement cost of $N$ ebits.  The classical communication cost would be $2N$ bits.  The above protocol requires one measurement, the entanglement between Alice and the $N$ parties is less than $\log_{2}(N+1)$, and the classical information transmitted is $\log_{2}(N+1)$ bits, because, since everyone receives the same message, there are only $N+1$ possible messages, and each of them has the same probability.

\section{Different states}
Suppose that instead of sending the same states to Bob and Charlie, Alice wants to send them different states, in particular, she wants to send $(1/\sqrt{2})(e^{i\theta_{1}} |0\rangle + e^{-i\theta_{1}}|1\rangle )$ to Bob and $(1/\sqrt{2})(e^{i\theta_{2}} |0\rangle + e^{-i\theta_{2}}|1\rangle )$ to Charlie.  She can do this by entangling a four-level system with two qubits, preparing the state
\begin{equation}
\label{difftemp}
\frac{1}{2} ( |0\rangle_{a} |00\rangle_{bc} + |1\rangle_{a}|01\rangle_{bc} + |2\rangle_{a}|10\rangle_{bc} + |3\rangle_{a}|11\rangle_{bc} ) .
\end{equation}
This resource state can be prepared by starting with two qubits in the state $|+x\rangle$ and the qudit in the state $|0\rangle$, that is, $|0\rangle_{a}|+x\rangle_{b}|+x\rangle_{c}$.  The qubits are connected to the qudit by Controlled-Shift gates, where, as before, the qubit is the control and the qudit the target.  The gate acts as (qubit state first) $|0\rangle |0\rangle \rightarrow |0\rangle |0\rangle$ and $|1\rangle |j\rangle \rightarrow |1\rangle |j+1\rangle$, where now the addition is modulo $4$.  Qubit $b$ is connected to qudit $a$ by two Controlled-Shift gates and qubit $c$ is connected to qudit $a$ by a single Controlled-Shift gate.  After passing through this circuit, the input state will be transformed into the state in Eq.\ (\ref{difftemp}).  An alternative is to replace the qudit by two qubits, $a1$ and $a2$, with the correspondence $|0\rangle_{a} \leftrightarrow |0\rangle_{a1}|0\rangle_{a2}$, $|1\rangle_{a} \leftrightarrow |0\rangle_{a1}|1\rangle_{a2}$, $|2\rangle_{a} \leftrightarrow |1\rangle_{a1}|0\rangle_{a2}$, and $|3\rangle_{a} \leftrightarrow |1\rangle_{a1}|1\rangle_{a2}$.  The state corresponding to that in Eq.\ (\ref{difftemp}) can then be produced with an input state $|0\rangle_{a1}|0\rangle_{a2}|+x\rangle_{b}|+x\rangle_{c}$ and a circuit with two C-NOT gates.  One of the C-NOT gates has qubit $b$ as the control and qubit $a1$ as the target, and the other has qubit $c$ as the control and qubit $a2$ as the target.

Note that the entanglement of the state in Eq.\ (\ref{difftemp}) between Alice, on the one hand, and Bob and Charlie on the other, is greater than in the case in which she sends the same states to Bob and Charlie, $3/2$ for the same state and $2$ for different states.  As before, she now sends qubit $b$ to Bob and qubit $c$ to Charlie.  Her operator $U_{a}$ now acts as follows
\begin{eqnarray}
U_{a}|0\rangle_{a} = e^{i(\theta_{1}+\theta_{2})} |0\rangle_{a} & U_{a}|1\rangle_{a}=e^{i(\theta_{1}-\theta_{2})} |1\rangle_{a} \nonumber \\
U_{a}|2\rangle_{a}=e^{-i(\theta_{1}-\theta_{2})} |2\rangle_{a} & U_{a}|3\rangle_{a} = e^{-i(\theta_{1}+\theta_{2})} |3\rangle_{a}  .
\end{eqnarray}
Alice then measures her system in the basis
\begin{equation} 
|u_{j}\rangle_{a} = \frac{1}{2} \sum_{k=0}^{3} e^{i\pi kj/2} |k\rangle_{a} ,
\end{equation}
for $j=0,1,2,3$.  She announces her result, and Bob applies the operator $U_{bj}$ to his qubit and Charlie applies $U_{cj}$ to his, where
\begin{eqnarray}
U_{bj} |0\rangle_{b} = |0\rangle_{b} & U_{bj} |1\rangle_{b} = (-1)^{j} |1\rangle_{b} \nonumber \\
U_{cj} |0\rangle_{c} = |0\rangle_{c}  & U_{cj} |1\rangle_{c} = e^{i\pi j/2} |1\rangle_{c} .
\end{eqnarray}
This will result in Bob and Charlie having the desired states. The only advantage in this case over standard teleportation is the reduced classical communication, in the sense that the message to Bob and Charlie is the same whereas in the case of teleportation there are, in general, two different messages.  The main point of this example is that it illustrates the fact that if one wants to send different messages to a number of receivers, the entanglement cost will be greater than if one wants to send them the same message.

\section{Qutrits}
Qutrits offer the possibility of Alice being able to control two parameters rather than one.  She is able to broadcast the equatorial state
\begin{equation}
\label{qutrit}
|\psi (\theta_{1},\theta_{2})\rangle = \frac{1}{\sqrt{3}} ( |0\rangle +e^{i\theta_{1}} |1\rangle + e^{i\theta_{2}}|2\rangle ) ,
\end{equation} 
to Bob and Charlie.  In order to do so, Alice starts with a $6$-level system entangled with two qutrits in the state
\begin{eqnarray}
|\Psi\rangle & = & \frac{1}{3} [ |0\rangle_{a} |00\rangle_{bc} + |1\rangle_{a} (|01\rangle_{bc} + |10\rangle_{bc}) \nonumber \\
& & +  |2\rangle_{a} |11\rangle_{bc} + |3\rangle_{a} (|02\rangle_{bc} + |20\rangle_{bc} ) \nonumber \\
& & + |4\rangle_{a} |22\rangle_{bc} + |5\rangle_{a} (|12\rangle_{bc} + |21\rangle_{bc}) ] . \label{eq:6-level}
\end{eqnarray}
Alice sends the qutrits to Bob and Charlie, and when she wants to broadcast  the states $|\psi (\theta_{1},\theta_{2})\rangle$ to them she applies the operator $U_{a}(\theta_{1},\theta_{2})$ to her system, where
\begin{eqnarray}
U_{a}(\theta_{1},\theta_{2})|0\rangle_{a} = |0\rangle_{a} & \hspace{2mm}& U_{a}(\theta_{1},\theta_{2})|1\rangle_{a} = e^{i\theta_{1}} |1\rangle_{a} \nonumber \\
U_{a}(\theta_{1},\theta_{2})|2\rangle_{a} = e^{2i\theta_{1}} |2\rangle_{a} & \hspace{2mm} & 
U_{a}(\theta_{1},\theta_{2})|3\rangle_{a} = e^{i\theta_{2}} |3\rangle_{a} \nonumber \\
U_{a}(\theta_{1},\theta_{2})|4\rangle_{a} = e^{2i\theta_{2}} |4\rangle_{a} & \hspace{2mm} &
U_{a}(\theta_{1},\theta_{2})|5\rangle_{a} = e^{i( \theta_{1}+\theta_{2}) } |5\rangle_{a} . \nonumber \\
\end{eqnarray} 
Next, she measures her system in the following basis 
\begin{eqnarray}
|u_{0}\rangle_{a} & = & \frac{1}{\sqrt{6}} \sum_{j=0}^{5} |j\rangle_{a} \nonumber \\
|u_{1}\rangle_{a} & = & \frac{1}{\sqrt{6}} \left[ \sum_{j=0}^{2} e^{2\pi ij/3} |j\rangle_{a} + e^{-2\pi i/3}|3\rangle_{a} + e^{2\pi i/3}|4\rangle_{a} \right.  \nonumber \\
&& \left. + |5\rangle_{a} \right] \nonumber \\
|u_{2}\rangle_{a} &= & \frac{1}{\sqrt{6}} \left[ \sum_{j=0}^{2} e^{-2\pi ij/3} |j\rangle_{a} + e^{2\pi i/3}|3\rangle_{a} + e^{-2\pi i/3}|4\rangle_{a} \right.  \nonumber \\
&& \left. + |5\rangle_{a} \right] \nonumber \\
|u_{3}\rangle_{a} & = & \frac{1}{\sqrt{6}}  \left[ \sum_{j=0}^{2} (-1)^{j}|j\rangle_{a} + \sum_{j=3}^{5} (-1)^{j}|j\rangle_{a} \right] \nonumber \\
|u_{4}\rangle_{a} & = & \frac{1}{\sqrt{6}} \left[ \sum_{j=0}^{2} (-1)^{j} e^{2\pi ij/3} |j\rangle_{a} - e^{-2\pi i/3}|3\rangle_{a} \right. \nonumber \\
 && \left. + e^{2\pi i/3}|4\rangle_{a} - |5\rangle_{a} \right] \nonumber \\
|u_{5}\rangle_{a} & = & \frac{1}{\sqrt{6}} \left[ \sum_{j=0}^{2} (-1)^{j} e^{-2\pi ij/3} |j\rangle_{a} - e^{2\pi i/3}|3\rangle_{a}  \right.  \nonumber \\
&& \left. + e^{-2\pi i/3}|4\rangle_{a} - |5\rangle_{a} \right]  ,  
\end{eqnarray}
and tells Bob and Charlie the result.  They can then apply correction operators to their qutrits, with the result that each will possess the quantum state $|\psi (\theta_{1},\theta_{2})\rangle$.  As an example, suppose that Alice obtained $|u_{2}\rangle_{a}$.  Bob and Charlie then have the state
\begin{equation}
\frac{1}{\sqrt{3}} (|0\rangle + e^{i\theta_{1}} e^{2\pi i/3} |1\rangle + e^{i\theta_{2}} e^{-2\pi i/3} |2\rangle ) .
\end{equation}
They can each correct this state by applying the operator $U=|0\rangle\langle 0| + e^{-2\pi i/3} |1\rangle\langle 1| + e^{2\pi i/3} |2\rangle\langle 2|$ to their qutrits.

This procedure will work with more general states than the one in Eq.~(\ref{qutrit}).  In analogy to the state in Eq.~(\ref{template}), we can introduce coefficients $\alpha$, $\beta$, and $\gamma$ into the state in Eq.~(\ref{qutrit}) and Alice can then send the state $|\psi (\theta_{1},\theta_{2})\rangle = ( \alpha |0\rangle + \beta e^{i\theta_{1}} |1\rangle + \gamma e^{i\theta_{2}}|2\rangle )$ to Bob and Charlie. We note that the state in Eq~(\ref{eq:6-level}) can be created in a way similar to the state used in the basic protocol by initializing the qudit in $|0\rangle$ and the two qutrits each in $(|0\rangle+|1\rangle+|2\rangle)/\sqrt{3}$, followed by Controlled-Shift gates by each qutrit. There are some differences. (a) The Controlled-Shift gate by the qutrit to the qudit is $|j\rangle|b\rangle\rightarrow|j\rangle|(b+j){\rm mod}\,N\rangle$. After the all Controlled-Shift gates, two parts in the resultant state are different from the desired forms: $|2\rangle_a (|02\rangle_{bc}+|20\rangle_{bc})$ and $|3\rangle_a(|12\rangle_{bc}+|21\rangle_{bc})$. (b) To turn the $a$ state in the first term to $|3\rangle_a$ and the second term to $|5\rangle_a$, one can apply further Controlled-Controlled-Shift gates (controlled by the two qutrits) so that only when the two qutrits are $|02\rangle_{bc}$ or $|20\rangle_{bc}$ the $a$ state is shifted by one unit and similarly by two units if the two qutrits are either $|12\rangle_{bc}$ or $|21\rangle_{bc}$.

\section{Applications}

What we have shown is that this protocol can be used to broadcast a class of qubit states to multiple parties.  The qubits could, for example, be used as inputs to quantum devices at different locations.  They could also be used as keys to unlock encoded quantum information.  Suppose Bob and Charlie each have a qubit in the state $|\psi (\theta )\rangle = (1/\sqrt{2}) (e^{i\theta} |0\rangle + e^{-i\theta}|1\rangle )$, which they have received from Alice.  Bob wants to send a qubit to Charlie in the state $(1/\sqrt{2}) (e^{i\phi} |0\rangle + e^{-i\phi}|1\rangle )$ and protect it from eavesdropping.  He takes his qubit in the state $|\psi (\theta )\rangle_{b}$, and rotates it about the $z$ axis by $\phi$ yielding the state $|\psi (\theta + \phi )\rangle_{b}$, and sends that to Charlie.  To any eavesdropper, who does not know the angle $\theta$, the state is
\begin{equation}
\frac{1}{\pi} \int_{0}^{\pi} d\theta\, |\psi (\theta + \phi )\rangle_{b}\langle \psi (\theta + \phi )| = (1/2)I_{b}  . 
\end{equation}
When Charlie receives Bob's qubit, he can combine it with his qubit yielding the state
\begin{eqnarray}
|\psi (\theta + \phi )\rangle_{b}|\psi (\theta )\rangle_{c} & = & \frac{1}{2}[e^{i(2\theta + \phi )} |00\rangle_{bc} + e^{i\phi}|01\rangle_{bc} \nonumber \\
& & + e^{-i\phi} |10\rangle_{bc} \nonumber \\
& & + e^{-i(2\theta + \phi )} |11\rangle_{bc}] .
\end{eqnarray}
If Charlie now measures this state with the projector $P=|01\rangle\langle 01| + |10\rangle\langle 10|$ and obtains the result $1$, which he does with a probability of $1/2$, and finally applies the unitary operator that takes $|01\rangle \rightarrow |01\rangle$ and $|10\rangle \rightarrow |11\rangle$, he will have the qubit Bob wanted to send him.

Quantum voting schemes have been proposed \cite{Ziman,Chefles,Bonanome,Horoshko,Pappa}, and the procedure developed here can also be used in a voting scheme to protect the votes from either an outside eavesdropper or from other voters.  Alice uses the procedure to send each of the voters the state $|\psi (\theta )\rangle$, and only Alice knows the value of $\theta$.  In order to vote each voter either does nothing or applies $\sigma_{z}$ to their qubit.  Note that $\sigma_{z} |\psi(\theta) \rangle = e^{-i\pi /2} |\psi (\theta - \pi /2)\rangle$, and $\langle \psi (\theta )| \psi (\theta -\pi /2)\rangle = 0$.  The qubits are then sent back to Alice, who, knowing $\theta$ can measure them in the correct basis to determine the votes.  Security can be increased by having Alice send several qubits to each voter, each with a different angle.  Alice announces the angles of several of the qubits, and the voters measure them in the corresponding bases to verify that what they received is the same as what Alice sent.  Any discrepancies would indicate the presence of an eavesdropper.  Each voter then performs the voting operations on his or her remaining qubits and sends them to Alice, who measures them to obtain the votes.  The qubits from each voter should yield the same vote, and if they do not, there has been tampering. 

If there are two receivers, it is possible to modify the procedure in order for Bob and Charlie to share a Bell state.  Specializing again to the case $\alpha = \beta = 1/\sqrt{2}$, Alice can measure the state in Eq.\ (\ref{template}) in the basis $(1/\sqrt{2}) ( |0\rangle \pm |1\rangle )$ and $|2\rangle$, and this will guarantee that Bob and Charlie share a Bell state, though which Bell state they share depends on the result of Alice's measurement.  If they wish, they can then use this state for teleportation.  For more than two receivers, this is no longer the case.  Measurements by Alice would result in the receivers sharing a multiparticle entangled state, and the utility of such a state for two-way communication is not clear.  Another possibility is the following.  Alice can apply a controlled unitary, $U_{c}$, to her qutrit.  In addition to her qutrit, she has a qubit, $a^{\prime}$, in the state $(|0\rangle + |1\rangle )/\sqrt{2}$, and the controlled unitary acts, for $j=0,1$, as
\begin{eqnarray}
U_{c}|j\rangle_{a^{\prime}}|0\rangle_{a} & = & e^{2i\theta_{j}} |j \rangle_{a^{\prime}}|0\rangle_{a}  \nonumber \\
U_{c}|j\rangle_{a^{\prime}}|1\rangle_{a} & = & e^{-2i\theta_{j}} |j \rangle_{a^{\prime}}|1\rangle_{a}  \nonumber \\
U_{c}|j\rangle_{a^{\prime}}|2\rangle_{a} & = &  |j \rangle_{a^{\prime}}|2\rangle_{a}  .
\end{eqnarray}
Measuring her qutrit and having Bob and Charlie apply the correction operations (note that these operations are independent of $\theta$) results in the state
\begin{equation}
\frac{1}{\sqrt{2}} (|0\rangle_{a^{\prime}} |\psi (\theta_{0})\rangle_{b} |\psi (\theta_{0})\rangle_{c} + |1\rangle_{a^{\prime}} |\psi (\theta_{1})\rangle_{b} |\psi (\theta_{1})\rangle_{c}) ,
\end{equation}
where $|\psi (\theta )\rangle = (1/\sqrt{2}) (e^{i\theta} |0\rangle + e^{-i\theta}|1\rangle)$.  If Alice now measures her qubit in the $|\pm x\rangle$ basis, Bob and Charlie will obtain the entangled state $ |\psi (\theta_{0})\rangle_{b} |\psi (\theta_{0})\rangle_{c} \pm  |\psi (\theta_{1})\rangle_{b} |\psi (\theta_{1})\rangle_{c}$ depending on her result.

\section{Conclusion}
We have presented a way of distributing a  restricted class of quantum states to multiple receivers.  The proposed protocol is a multipartite  version of remote state preparation, and it uses less entanglement than making use of teleportation to accomplish the same result.  It provides an alternative to teleportation for some tasks in quantum networks.  A number of examples were presented.  These results should facilitate communication in quantum networks.

\acknowledgments
The research of MH, JB, and T-CW was supported by NSF grant FET-2106447. The research of MH and JB was also sponsored by the DEVCOM Army Research Laboratory and was accomplished under Cooperative Agreement Number W911NF-20-2-0097. We would like to thank Himanshu Gupta for useful comments.

\end{document}